# Width-independent and Robust Multimode Interference Waveguides Based on Anomalous Bulk States

*Lei Liu[1], Xiujuan Zhang[1, *], Ming-Hui Lu[1, 2, 3], and Yan-Feng Chen[1, 3]*

[1]National Laboratory of Solid State Microstructures and Department of Materials Science and Engineering, Nanjing University, Nanjing 210093, China.

[2]Jiangsu Key Laboratory of Artificial Functional Materials, Nanjing 210093, China.

[3]Collaborative Innovation Center of Advanced Microstructures, Nanjing University, Nanjing 210093, China.

*Corresponding authors. E-mail: xiujuanzhang@nju.edu.cn

**Multimode interference (MMI) is a fundamental physical principle that plays a crucial role in modern communication technologies for wave splitting, filtering, switching and multiplexing. Typically, the generation of multimodes is highly dependent on the waveguide's cross-section, particularly its width, by which the mode profiles and the interference patterns can be severely affected, leading to unstable MMI performance. Here, we realize width-independent and robust MMI waveguides. Our principle is based on the unique properties of multilayer graphene lattices. By properly modulating the boundary potential, this Dirac-type material supports anomalous bulk states with uniform wavefunctions independent of the sample size. Benefited from such an anomaly, the bulk states in a waveguide formed by multiple layers of graphene ribbons exhibit width-independent MMI. Enabled by this intriguing characterisitc, we construct $2 \times 2$ MMI waveguides using bilayer photonic and phononic crystals with graphene lattices. By precisely modulating their boundaries, the anomalous bulk states and correspondingly the width-independent MMI are achieved. Our experimental measurements show that the input wave energy travelling through the MMI waveguide can be split into two outputs with a frequency-tunable ratio, and due to the width-independent characteristic, the splitter is robust to geometric perturbations. We further demonstrate stable MMI across multiple interconnects with stepped widths, allowing for high power capacity while maintaining high coupling efficiency. Our approach successfully decouples MMI performance from waveguide width, creating lateral degrees of freedom that enable flexibly scalable and robust photonic, phononic, and electronic integrated circuits for versatile MMI applications.**

# 1. Introduction

Multimode interference (MMI) occurs when waves entering a waveguide excite multiple propagating modes, producing interference patterns.[1] These interference patterns form new wavefronts which can be harnessed to realize wave guiding, routing, dividing, coupling, and multiplexing. This mechanism is fundamental for spatial wave manipulations and is widely used in applications like optical fiber communications,[2–8] and on-chip optics,[1,9–15] enabling photonic integrated circuits,[9,10] spectrometry,[6,16] imaging,[5,8] deep learning,[13] and quantum computing.[9,11,14] An MMI device consists of three main components: inputs, MMI waveguide and outputs. The inputs and outputs are often narrow waveguides in order to connect with laser sources or for chip-to-chip interconnects. The width of the MMI waveguide, on the other hand, is strictly designed for specific wave manipulation tasks and typically mismatched with the inputs and outputs. This can significantly affect the mode profiles and interference patterns, leading to insertion loss due to reflections, particularly when waves pass across multiple interconnects with different widths. Consequently, MMI devices are highly sensitive to the waveguide width, which can limit power capacity and coupling efficiency with other systems. Upon geometric perturbations or imperfections, the MMI performance also can become unstable. To address these challenges, the key is to achieve independent and robust control over both MMI and the waveguide width.

Over the past two decades, metamaterials have attracted tremendous attention due to their ability to regulate material dispersions beyond traditional limits, enabling unprecedented wave phenomena such as zero-refraction,[17–19] negative-refraction,[20–22] and material anisotropy.[23,24] Facilitated by versatile artificial designs, metamaterials have particularly advanced the development of topological materials which are well-known for their topological edge states featuring robust wave transports.[25,26] Such robust wave transports are usually enabled by specific lattice symmetries and protected by nontrivial topological invariants.[27] Recently, it has been counterintuitively found that in topologically trivial lattices, robust wave transports can also be realized by engineering boundaries.[28–35] Different from the topological edge states, these boundary-induced modes are bulk states which can extend across the entire footprint of the bulk region. This unique property offers new opportunities for width-independent wave manipulations.

Here, we explore the mechanism of boundary engineering to realize anomalous bulk states in a type of Dirac material with graphene lattice. By vertically stacking the designed materials with proper boundary modulations, a ribbon-like waveguide is constructed, which supports multimodes featuring uniform wavefunctions extended across the entire bulk region. We theoretically demonstrate the width independence of the interference between these multimodes based on a tight-binding method. For the verification, standard $2 \times 2$ MMI splitters are designed using bilayer photonic and phononic crystals composed of graphene lattices. By imposing hard boundaries, we achieve anomalous bulk states with even and odd field distributions. Their interference is shown to be width-independent, agreed with the theory.

Experimental measurements confirm that input waves can be split into the output channels, with the splitting ratio continuously regulated by excitation frequencies while maintaining a stable MMI performance even in the presence of geometric perturbations. Based on these intriguing characteristics, we further design MMI devices with multiple interconnects, enabling robust wave transports that can sustain steady power flow and high coupling efficiency despite the abrupt changes in waveguide width. Our principle based on the anomalous bulk states advances the practical applications of robust wave transports over a large footprint. The realized width-independent and robust MMI holds great promise for integrated circuits with flexible scalability, high power capacity and high coupling efficiency, applicable to devices such as splitters, filters, couplers, logic gates, and multiplexers.

## 2. Results

### 2.1. MMI waveguides and their width-independent characteristics

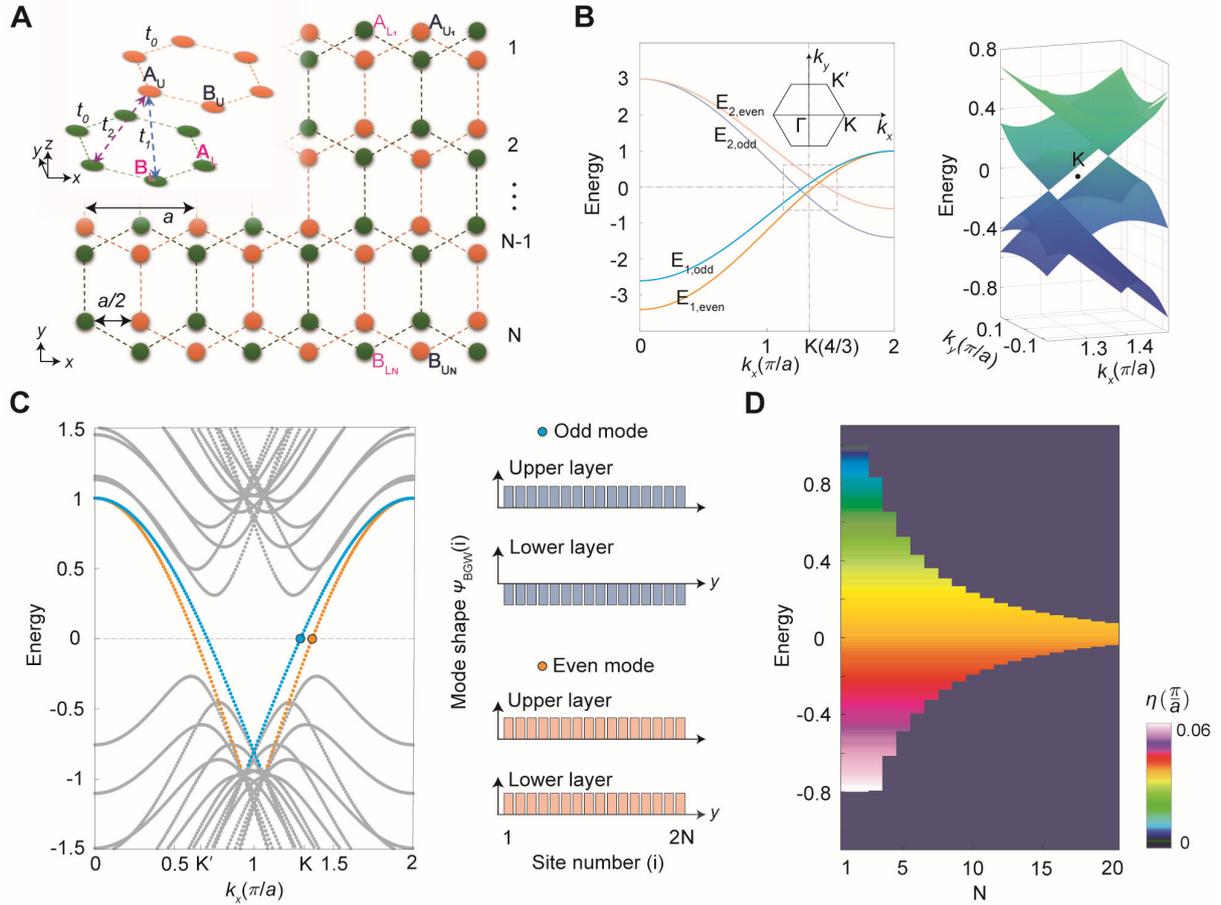

**Figure 1.** Multimode anomalous bulk states in a bilayer graphene waveguide (BGW). A) Schematics of the waveguide from the side and top views. The green and yellow symbols constitute the lower and upper graphene lattices, respectively. A and B mark the lattice sites. Waveguide boundaries are opened in the $y$-direction consisting of zigzag edges, with $N$ measuring the waveguide width as $W = N\sqrt{3}a/2$. B) Left panel: band structures of the bulk bilayer graphene along the $k_x$ direction. Right panel: energy surfaces around the $K$ point exhibiting two Dirac cones. C) Band structures of the BGW with width $W = 4\sqrt{3}a$ ($N = 8$)

and boundary potential taken as $t_0$. The even and odd anomalous bulk states are denoted by the orange and blue dots, respectively, with normalized wavefunctions shown in the right panel. D) Interference factor $\eta$ (color map) as functions of the waveguide width and the excitation frequency (energy). The dark regions represent areas where other regular higher-order bulk states exist. The coupling coefficients are taken as $t_0 = -1$, $t_1 = -0.2$ and $t_2 = -0.2$.

We begin with a hexagonal graphene lattice which supports a Dirac point at the high-symmetric point K, belonging to the Dirac materials.[36] By vertically stacking this type of material, multimodes can be realized. Here, without loss of generality, we focus on a bilayer design (see Section 1 of Supporting Information for analyses on multi-layer designs). As illustrated in **Figure 1**A, the upper layer slides half a lattice relative to the lower layer along the $x$-direction. This is for the purpose of introducing interlayer couplings, which, as will be discussed in the following, control the interference between the mutimodes. The coupling coefficients are denoted by $t_1$ and $t_2$. Within each monolayer, the intralayer coupling is characterized by $t_0$ (only nearest coupling is considered). The Hamiltonian for such a model comprises an intralayer part $H_{\text{intra}}(\boldsymbol{k})$ and an interlayer part $H_{\text{inter}}(\boldsymbol{k})$, and reads

$$H(\boldsymbol{k}) = H_{\text{intra}}(\boldsymbol{k}) + H_{\text{inter}}(\boldsymbol{k}) = \begin{bmatrix} h(\boldsymbol{k}) & 0_{2\times2} \\ 0_{2\times2} & h(\boldsymbol{k}) \end{bmatrix} + \begin{bmatrix} 0_{2\times2} & K(\boldsymbol{k}) \\ K(\boldsymbol{k})^{\dagger} & 0_{2\times2} \end{bmatrix}, \qquad (1)$$

where $h(\boldsymbol{k}) = (2t_0 \cos\frac{k_x}{2} + t_0 \cos\frac{\sqrt{3}k_y}{2})\sigma_x - t_0\sin\frac{\sqrt{3}k_y}{2}\sigma_y$ represents the Hamiltonian of the monolayer graphene, and $K(\boldsymbol{k}) = t_2 \cos\frac{k_x}{2}\sigma_0 + t_1\sigma_x$ the interlayer coupling matrix ($\sigma_0$ is the identity matrix, and $\sigma_x$ and $\sigma_y$ are the Pauli matrices). Here, the lattice constant is taken as $a = 1$, $\boldsymbol{k}(k_x, k_y)$ denotes the wave vector, and † indicates the complex conjugate transpose. $H$ corresponds to eigenstate $|\psi\rangle = (\phi_{A_L}, \phi_{B_L}, \phi_{A_U}, \phi_{B_U})^T$, where $\phi_{ij}$ with $i = A, B$ and $j = L, U$ represents wavefunction on site $i$ in the $j$-layer (L for the lower layer and U for the upper layer). Consider wave propagating along the $x$-direction, obeying the eigenequation $H|\psi\rangle = E|\psi\rangle$ with $k_y = 0$. Solving the eigenequation yields the following eigenenergies and eigenstates

$$E_{1,\text{even}} = (t_0 + t_1) + (2t_0 + t_2)\cos\frac{k_x}{2}, |\psi\rangle_{1,\text{even}} = \frac{1}{2}[1 \quad 1 \quad 1 \quad 1]^T, \qquad (2a)$$

$$E_{1,\text{odd}} = (t_0 - t_1) + (2t_0 - t_2)\cos\frac{k_x}{2}, |\psi\rangle_{1,\text{odd}} = \frac{1}{2}[-1 \quad -1 \quad 1 \quad 1]^T, \qquad (2b)$$

$$E_{2,\text{odd}} = -(t_0 - t_1) - (2t_0 + t_2)\cos\frac{k_x}{2}, |\psi\rangle_{2,\text{odd}} = \frac{1}{2}[1 \quad -1 \quad -1 \quad 1]^T, \qquad (2c)$$

$$E_{2,\text{even}} = -(t_0 + t_1) - (2t_0 - t_2)\cos\frac{k_x}{2}, |\psi\rangle_{2,\text{even}} = \frac{1}{2}[-1 \quad 1 \quad -1 \quad 1]^T. \qquad (2d)$$

There are four energy bands in total, with even and odd indicating the parity of the eigenstates with respect to the lower and upper layers.

The key technique to the boundary engineering is to choose a proper boundary potential such that it matches with the average energy potential of the bulk, as if the system were boundaryless. As an example, we consider the coupling coefficients of $t_0 = -1$, $t_1 = -0.2$ and $t_2 = -0.2$ (this parameter selection is to be comparable with the physical material design as presented in the following sections; see further details in Section 2 of Supporting Information). The energy bands are shown in Figure 1B, which exhibit two Dirac points near the K point. The deviation from the K point is due to the interlayer coupling originated from sliding the two graphene layers. Nonetheless, the eigenstates in the bilayer structure are not affected and inherent the same characteristics from the monolayer graphene, as reflected from the eigenstate distributions in Equation (2). It indicates that for each energy band, the eigenstates share the same field distribution, regardless of variations on $k_x$ or the coupling coefficients. This is a unique property in graphene lattices, which is essential to the boundary engineering. Otherwise, for eigenstates strongly dependent on wave vectors or coupling coefficients, it is unlikely to construct a physical boundary that matches with the bulk potential.

For our case, we write out the Hamiltonian of the bilayer graphene waveguide (BGW) with boundaries in the $y$-direction, yielding $H_{BGW}(k_x, E_{edge})|\psi_{BGW}\rangle = E_{BGW}|\psi_{BGW}\rangle$ under the basis of $|\psi_{BGW}\rangle = (\phi_{A_{L_1}}, \phi_{B_{L_1}}, \phi_{A_{U_1}}, \phi_{B_{U_1}}, \cdots, \phi_{A_{L_N}}, \phi_{B_{L_N}}, \phi_{A_{U_N}}, \phi_{B_{U_N}})^T$ [see Section 3 of Supporting Information for the full expression of $H_{BGW}$, which can be further decomposed into the lower- and upper-layer components as $|\psi_L\rangle = (\phi_{A_{L_1}}, \phi_{B_{L_1}}, \cdots, \phi_{A_{L_N}}, \phi_{B_{L_N}})^T$ and $|\psi_U\rangle = (\phi_{A_{U_1}}, \phi_{B_{U_1}}, \cdots, \phi_{A_{U_N}}, \phi_{B_{U_N}})^T$]. Here, N denotes the cell number in the $y$-direction. $E_{edge}$ represents the boundary potential to be determined by $|\psi_{BGW}\rangle$. This constitutes a general inverse problem that can be solved numerically. In our case, fortunately, based on the characteristics of the bulk eigenstates $|\psi\rangle$ in Equation (2), the average bulk energy potential can be directly deduced as $\pm t_0$, corresponding to $|\psi_{BGW}\rangle$ being N-duplicates of $|\psi\rangle$. The boundary potential set as $t_0$ or $-t_0$ determines the survival of $|\psi\rangle_1$ or $|\psi\rangle_2$ group of bulk eigenstates, respectively, independent of N. Consider $E_{edge} = t_0$ and plot the spectral of $E_{BGW}$, as shown in Figure 1C. Two energy bands appear, with dispersions indeed resembling that of the $E_1$ group in Figure 1B. The wavefunctions (the right panels of Figure 1C) provide more evidence, which are uniformly distributed along the $y$-direction and share the same parity with the bulk eigenstates $|\psi\rangle_{1,even}$ and $|\psi\rangle_{1,odd}$. Naturally, if $E_{edge}$ is set as $-t_0$, the $E_2$ (and $|\psi\rangle_2$) group will be selected and survive as the anomalous bulk states (see details in Section 4 of Supporting Information, which also discusses other boundary potentials away from the average bulk potential).

Under excitations along the $x$-direction, the even and odd anomalous bulk states are interfered and superposed, leading to MMI fields as

$$|\Omega_{MMI}\rangle = C_1 e^{ik_{even}x}|\psi_{BGW}\rangle_{even} + C_2 e^{ik_{odd}x}|\psi_{BGW}\rangle_{odd}, \qquad (3)$$

where $C_1$ and $C_2$ are the excitation coefficients and $k_{\text{even}}$ and $k_{\text{odd}}$ are the wavevectors corresponding to the two anomalous bulk states. The wavefunction density distribution is accordingly obtained as

$$|\Omega_{\text{MMI}}|^2 = \begin{cases} |\psi_{\text{U}}|^2 \cos^2 \eta x & \text{for upper layer} \\ |\psi_{\text{L}}|^2 \sin^2 \eta x & \text{for lower layer} \end{cases}, \text{ with } \eta = \left| \frac{k_{\text{even}} - k_{\text{odd}}}{2} \right|. \qquad (4)$$

Here, unbiased excitations are considered with $C_1 = C_2$, yielding $|C_1|^2 + |C_2|^2 = 1$ under the energy conservation. Note that Equation (4) describes precisely the characteristics of two-mode interference.[37] The interference factor $\eta$ can be rewritten into

$$\eta = \left| \frac{4\sqrt{3}(t_2 E_0 + 2t_0 t_1 - t_0 t_2)}{3(4t_0{}^2 - t_2{}^2)} \right|, \qquad (5)$$

where the Taylor expansion is used around the K point, up to the linear order $O(k)$. $E_0 (= \hbar\omega)$ represents the excitation frequency. From Equation (5), it can be seen that $\eta$ is independent of N, suggesting that the interference between the anomalous bulk states is indeed width independent. Furthermore, the interference pattern can be tuned by the coupling coefficients and, more importantly, by the frequency. These characteristics are precisely reflected in Figure 1D where $\eta$ is presented as functions of waveguide width and excitation frequency. Note that there are two dark regions accommodating other regular higher-order bulk states, whose existence demarcates a band gap for the MMI. As N increases, the higher-order bulk states above and below the zero energy gradually approach to each other, by which the band gap shrinks (see Section 5 of Supporting Information for discussions on the origin of such band gap due to the finite-size effect and its relationship with N). This can narrow the MMI window as shown in Figure 1D. Nonetheless, the MMI with width-independent property is unaffected and a full interference cycle is always achievable by tuning the coupling coefficients while maintaining a compactness along the $x$-direction.

## 2.2. A 2 × 2 MMI splitter based on bilayer photonic and phononic crystals

Based on the principle of the width-independent MMI, we design a standard 2 × 2 MMI splitter. As illustrated in **Figure 2**A, the splitter consists of two inputs and two outputs, connected by the MMI waveguide made of bilayer phononic crystals. Each layer is composed of a hexagonal graphene lattice of cylindrical rods with a diameter of $d = 7.0$ mm and a thickness of $h = 10$ mm. The lattice constant is $a = 12$ mm. The two layers are relatively shifted by half a lattice constant $a/2$ along the $x$-direction. Hard boundaries are imposed in both the $z$- and $y$-directions, to form a waveguide. The $z$-boundaries are used to confine waves, while the $y$-boundaries, on the other hand, are for the purpose of boundary engineering. Similar to the theoretical analyses, we first calculate the band structures for the bilayer phononic crystals along the $k_x$ direction, as shown in Figure 2B. As predicted, two Dirac cones emerge near the K points, indicated by the colored bands, while the gray ones represent the higher-order bulk states. The right panels of Figure 2B present the pressure filed $P$ distributions of the four eigenstates marked in the left panel. The two groups of eigenstates again exhibit even and odd parities with respect to the upper and lower layers, which can be seen from the sign changes of $P$. This is consistent with the theory.

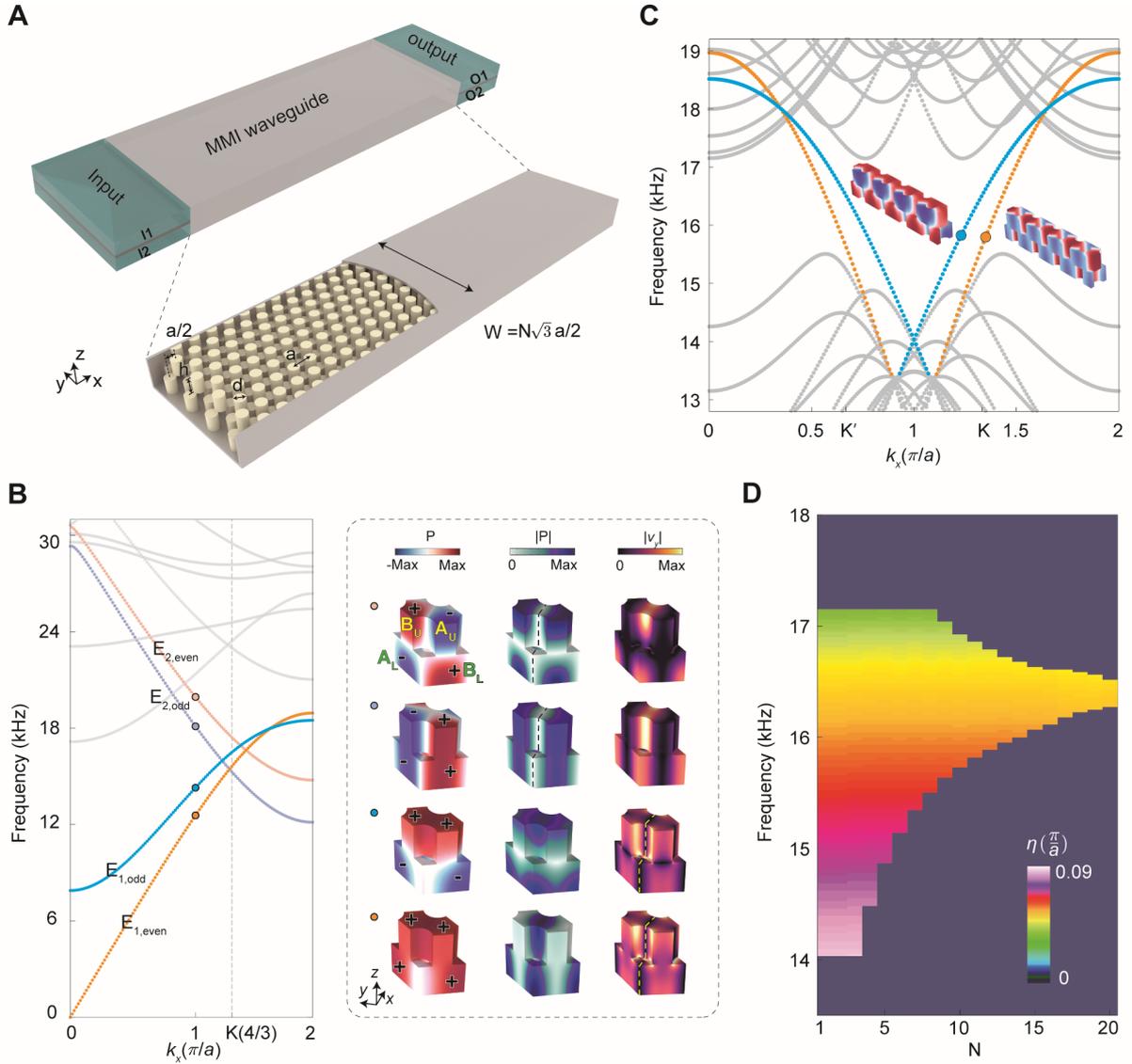

**Figure 2.** A $2 \times 2$ MMI splitter based on bilayer phononic crystals. A) Schematic of the splitter, consisting of two inputs, the MMI waveguide, and two outputs. B) Left panel: bulk band structures along the $k_x$ direction. Right panel: acoustic field and amplitude distributions, $P$, $|P|$ and $|v_y|$, with the parities of the pressure fields indicated by + and − signs. The black and yellow dashed lines mark the planes with $|P| = 0$ and $|v_y| = 0$, respectively. C) Band structures of the waveguide supercell, with $N = 8$. Insets: pressure fields of the odd (blue dot) and even (orange dot) anomalous bulk states. Their interference factor $\eta$ is presented in (D).

To engineer the $y$-boundaries, we further present the amplitude distributions $|P|$ and $|v_y|$ (with $v_y$ the velocity fields along the $y$-direction). By closely examining $|P|$ and $|v_y|$, it is found that the eigenstates in the $E_1$ and $E_2$ groups exhibit unique characteristics. For the $E_1$ group, $|v_y|$ is uniformly distributed on a plane in the $z$-direction, where $|v_y| = 0$, as marked by the yellow dashed lines. For the $E_2$ group, the same plane is also a zero-plane, but accommodating zero $|P|$ (as marked by the black dashed lines). If we set the boundaries precisely at these zero-planes, the waves cannot differentiate whether these are boundaries or wave nodes in the bulk, effectively matching the boundary potential with the bulk. In this way, the boundary engineering condition is satisfied and the system behaves as boundaryless. In other words, the $|v_y| = 0$ plane effectively produces $E_{edge} = t_0$ while the $|P| = 0$ plane produces $E_{edge} = -t_0$. In practice, $|v_y| = 0$ corresponds to physical hard boundaries while $|P| = 0$ to soft boundaries.[38] Here, we consider the most common boundaries for waveguides, i.e., the hard boundaries (see analyses for other boundary engineering including soft boundaries in Section 6 of Supporting Information). The hard boundaries naturally select the eigenstates in the $E_1$ group as the anomalous bulk states, as shown by the band structures of the waveguide supercell with N = 8 in Figure 2C. The pressure filed distributions further demonstrate their uniform wavefunctions, as well as the parity properties. The interference factor $\eta$ between these multimodes is shown in Figure 2D, which again exhibits width-independent and frequency-tunable characteristics.

We argue that this principle of boundary engineering is universal and further demonstrate the width-independent MMI in a silicon waveguide made of bilayer photonic crystals, as detailed in Section 7 of Supporting Information. It is shown that with appropriate boundary engineering, the electromagnetic waves also exhibit width-independent transitions across a broad regime of optical/fiber communication frequencies.

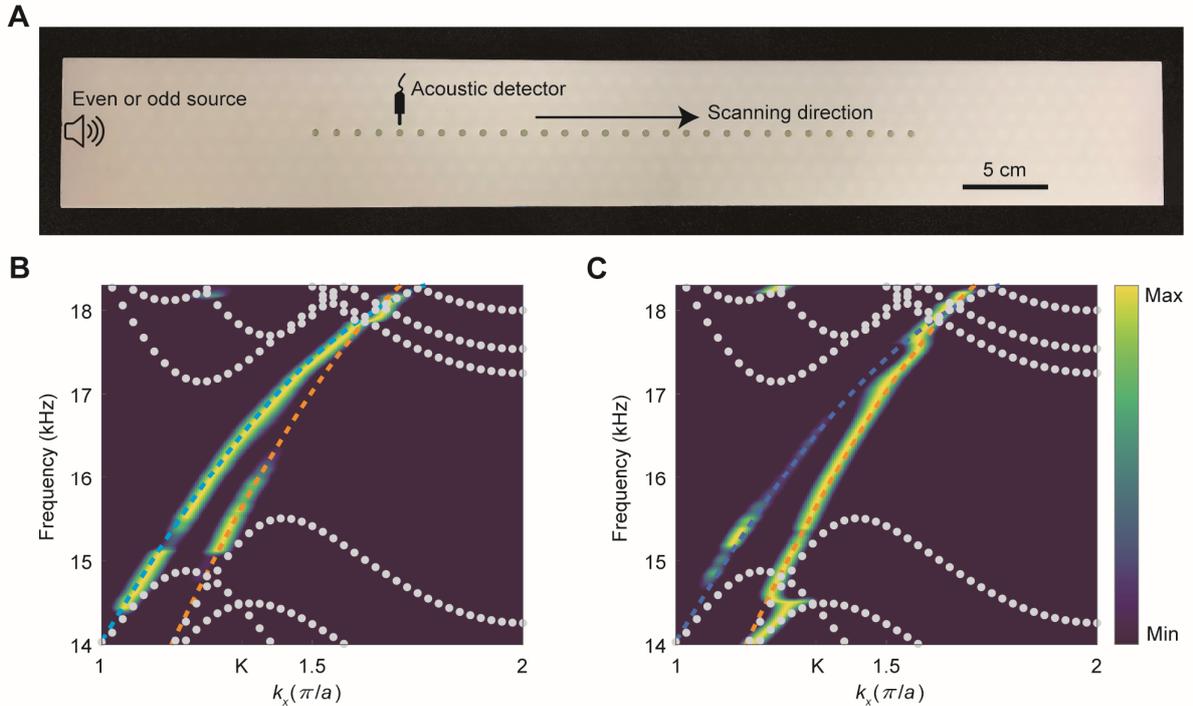

**Figure 3.** Experimental measurements of the anomalous bulk states. A) A photo of the sample and the experimental setup. The measured dispersions (color maps) are presented in (B), for the odd states and in (C), for the even states, compared to the simulations (with the blue and orange dashed lines indicating the odd and even states, respectively, and the grey dots denoting the other regular bulk states).

To experimentally measure the anomalous bulk states, we fabricate a waveguide sample of the bilayer phononic crystals as shown in **Figure 3**A, using 3D printing technology with resin (an acoustically hard material). To specifically excite the even (or odd) states, two loud speakers with phase difference of 0 (or $\pi$) are placed in the upper and lower inputs to create an even (or odd) source. The pressure amplitudes and phases are measured via an acoustic detector along the $x$-direction through holes drilled on top of the sample. The collected real-space data are then Fourier transformed to obtain the dispersions in the momentum space, as shown in Figure 3B and 3C, respectively for the odd and even states. Simulations as that in Figure 2C are also presented as comparisons. It is observed that the measured data are in high agreements with the simulations, especially in the band gap where the measurements unambiguously demonstrate the existence of even and odd anomalous bulk states. A small discrepancy appears out of the band gap, around 15.3 kHz, which is likely due to mixing with the nearby regular bulk states.

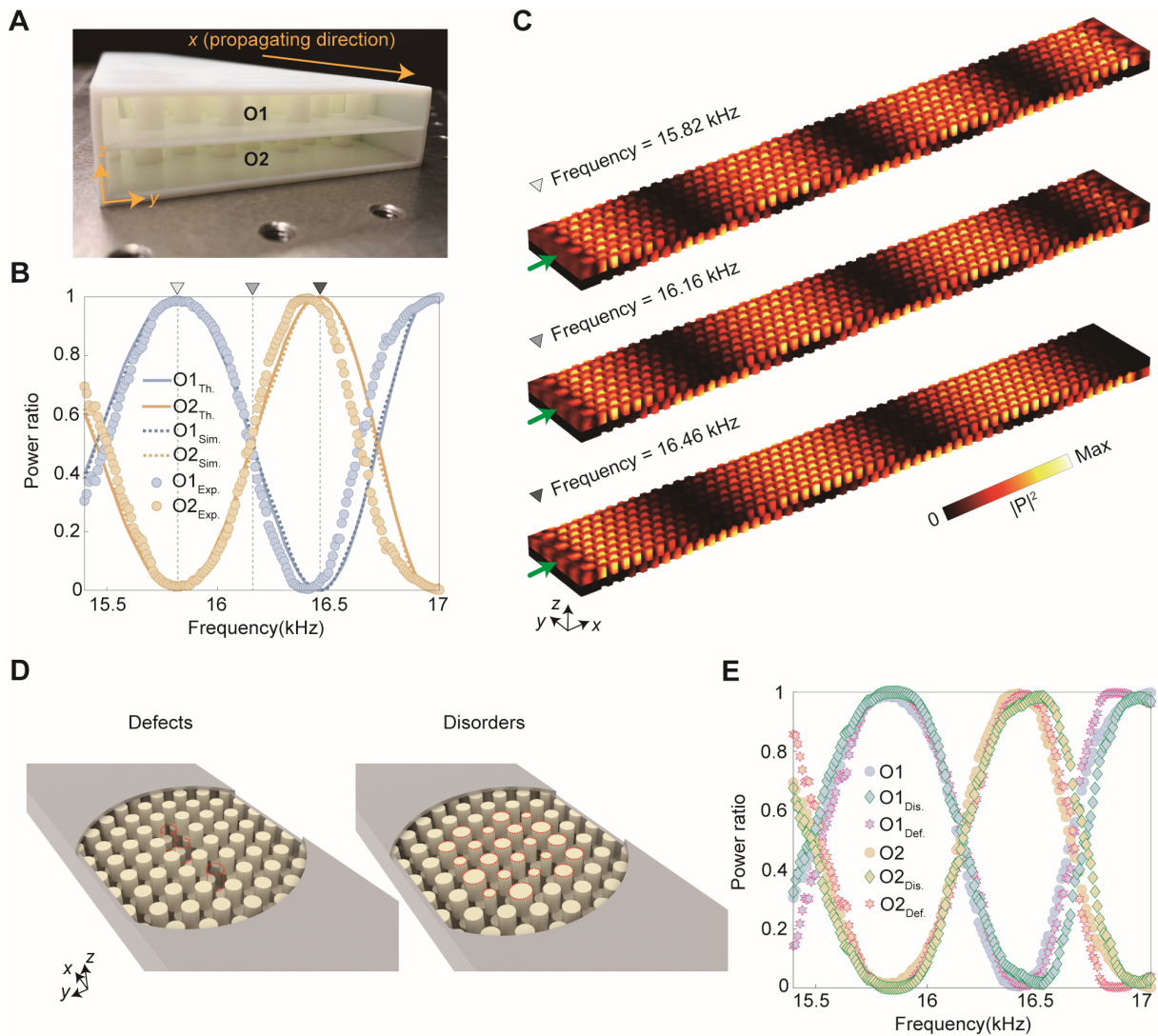

**Figure 4.** Frequency-tunable and robust MMI power splitting. A) A photo of the MMI splitter. The incident power comes from the upper-layer input I1, propagates through the MMI waveguide, and eventually splits into the outputs O1 and O2. B) Output power ratios in O1 (blue) and O2 (orange) ports as a function of frequencies. Solid lines, dotted lines, and dots represent theoretical, simulated, and experimental results, respectively. C) Simulated intensity distributions $|P|^2$ in the MMI splitter with different incident frequencies. The green arrows mark the input. D) Illustrations of defects and disorders. E) Measured output power ratios in O1 and O2, with and without defects and disorders.

As the main functionality, the MMI waveguide can spilt the input power into multiple outputs. The splitting power ratio is often specifically determined based on the individual waveguide size and typically has an undefined (or weak) correlation with the frequencies.[39,40] Here, we demonstrate that based on our principle, arbitrary power ratio from 50:50 to 100:0 (also to the reciprocal ratio of 0:100) can be obtained and more remarkably this can be continuously regulated by changing the excitation frequencies. Consider the incident power from the input I1 and measure the split power at the outputs O1 and O2, as $|P_1|^2$ and $|P_2|^2$, respectively (see illustrations in **Figure 4**A). The power ratio can be accordingly obtained as $\frac{|P_{1\,or\,2}|^2}{|P_1|^2+|P_2|^2}$. Different from the measurements in Figure 3, here, the source is unbiased, hence exciting both the even and odd anomalous bulk states, yielding MMI fields and power distributions as described in Equation (3) and (4). The measured power ratio is presented in Figure 4B compared with both the theoretical calculations following Equation (4) (with $x = 48.5a$ being the length of the MMI waveguide) and the numerical simulations. The three sets of data exhibit excellent agreements with each other. As expected, different output power ratios are indeed achieved at different frequencies, obeying a continuous regulation. Figure 4C visualizes the power splitting into $O1: O2$ with the ratio of 100:0, 50:50 and 0:100, at different frequencies. Such a frequency-tunable behavior is exactly originated from the linear dependence of $\eta$ on frequencies, as revealed by Equation (5).

The MMI splitter based on the multimode anomalous bulk states is also robust to local geometric perturbations. Such robustness is firstly attributed to the fact that the anomalous bulk states originate from the Dirac dispersions, which are protected by the time and spatial inversion symmetries (see Section 2 of Supporting Information). In addition, the anomalous bulk states around the K point have a large momentum mismatch with their time-reversal partners around the K′ point (with $\Delta k \approx \frac{2\pi}{3a}$), which can effectively suppress the backward scattering. For demonstrations, we deliberately introduce wavelength-scale defects and disorders into the waveguide, by removing several scattering rods and by changing the diameters of some rods with a randomness in the range of $[0.7d, 1.3d]$, as illustrated in Figure 4D. The power ratios of O1 and O2 ports with defects and disorders are again measured, whose results are shown in Figure 4E, compared to the data without defects and disorders. Again, excellent agreements are observed, validating the robustness. As a comparison, in Section 8 of Supporting Information, we conduct simulations on MMI in conventional phononic line-defect waveguides, which is shown to be rather vulnerable to defects and disorders.

## 2.3. Stable MMI and high coupling efficiency across multiple interconnects

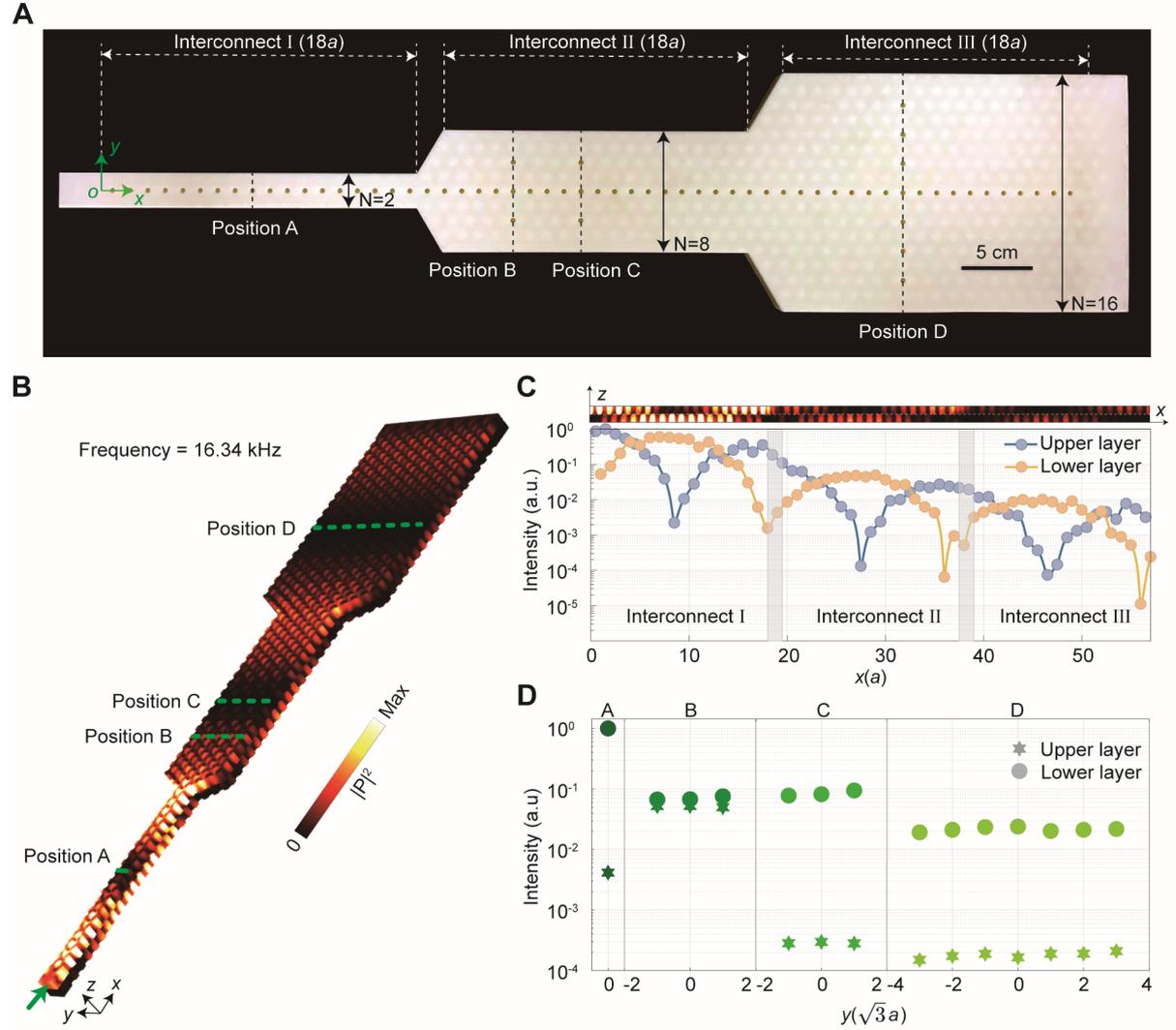

**Figure 5.** MMI in multiple interconnects with stepped widths. A) A photo of the MMI device consisting of three interconnects with different widths. B) Simulated intensity distributions $|P|^2$ of the incident power from the I1 port. C) Measured $|P|^2$ along the middle line ($y = 0$). The grey shadings indicate the intermediates between neighboring interconnects. The top panel shows the simulated $|P|^2$ in a cut-plane plot at $y = 0$. D) Measured $|P|^2$ along the $y$-direction at positions A, B, C, and D [as marked in (A) and (B)].

Power transition across multiple interconnects with different widths while maintaining stable MMI and a high coupling efficiency is always challenging due to reflections. Here, we show that owing to the width-independent and robust characteristics, the multimode anomalous bulk states across multiple interconnects can keep a steady interference pattern and enable high power coupling efficiency, despite abrupt width changes. As illustrated in **Figure 5**A, two additional interconnects (with N = 2 and N = 16) are inserted, in front of and behind the original MMI waveguide (with N = 8). For the sake of clarity, the length of the interconnects is carefully designed to cover approximately one interference period, i.e., $\eta x$ covers $[0, \pi]$. From Figure 2D, it is found that $\eta$ takes a value of $0.054\frac{\pi}{a}$ at the frequency of 16.34 kHz, which gives the interference period rounding to $x = 18a$. It is accordingly set as the length of the interconnects.

We still consider the incident power from the I1 port. Figure 5B presents the simulation on the power transition across the three interconnects with different widths, which, as expected, indeed exhibits steady MMI effect. The power change in each interconnect follows $\cos^2\eta x$ and $\sin^2\eta x$, respectively for the upper and lower layers, consistent with the theoretical descriptions in Equation (4). It is also noticed that the power density decreases step-wisely across each interconnect, while maintaining a unity total power (see Section 9 of Supporting Information for details). This is because for a given input power, the wider the waveguide is, the smaller the power density is, and vice versa. This interesting phenomenon has a significant implication that our device, without affecting the MMI, can freely squeeze or expand the same amount of power regardless of the waveguide width, which may inspire novel waveguide designs with high power capacity and high coupling efficiency.

Experimentally, we measure the power intensity along the middle line ($y = 0$), as shown in Figure 5C. Three groups of repeated curves are observed, each accommodating an interference period approximately $18a$, matching the theoretical value. Moreover, the power density exhibits apparent step-wise decreasing, also aligning with the simulation as shown in the top panel (or in Figure 5B). In Section 10 of Supporting Information, we also present the simulation and the experimental measurements on a reversed power transition (i.e., the power is inputted from the O1 port and the I1 and I2 ports become the outputs). The results exhibit again steady MMI, only that the power transitioning from a wide waveguide is now squeezed into a narrow one. To verify the uniform intensity in the width direction, further measurements are conducted along the $y$-direction at specific $x$-positions (as marked in Figure 5A and 5B). Positions A, C, and D are set to be the middle of each interconnect corresponding to the power ratio of $0:100$ between the upper and lower layers, while position B is chosen for the power ratio of $50:50$. The measured intensity is shown in Figure 5D, which is seen to be uniform along the $y$-direction for both the upper and lower layers at all positions. Particularly, the intensity at positions A, C, and D is concentrated in the lower layer, indicating a $0:100$ power ratio. On the other hand, at position B, the power is evenly spilt into the upper and lower layers with a $50:50$ power ratio, again exhibiting excellent agreements with the theoretical predictions.

## 3. Conclusion

We have achieved multimode anomalous bulk states via boundary engineering. This unique principle guarantees a boundaryless system, hence allowing uniform wave distributions across a large footprint of the entire bulk region, which are both width-independent and robust to geometric perturbations. Owing to these characteristics, the anomalous bulk states are explored in MMI devices, enabling robust power splitting with ratios that can be continuously regulated by the excitation frequencies. The width-independent property further facilitates power transitions across multiple MMI interconnects with abrupt width changes, maintaining steady power flow and high coupling efficiency. These MMI devices are experimentally realized based on bilayer phononic crystal waveguides with hard boundaries and are also proposed and simulated in silicon waveguides made of bilayer photonic crystals at optical and fiber communication frequencies. Our approach offers a novel mechanism for MMI beyond the traditional designs that heavily depend on the waveguide widths and are vulnerable to the geometric perturbations. Targeting on the vast MMI functionalities in optical communications and on-chip photonics/phononics, the relaxation on the waveguide widths further creates a lateral degree of freedom for chips and fibers with flexible scalability and high coupling/transmission efficiency. Furthermore, our theory is based on tight-binding models, rendering the principle universal not only in bosonic systems but also in fermionic systems where robust current control over a scalable footprint is crucial, such as in graphene electronics.[41,42]

## 4. Methods

*Simulations*

All simulations in this work are carried out using the acoustics module of COMSOL Multiphysics. The speed of sound and the air density used are 343 m/s and 1.21 kg/m$^3$, respectively. In Figure 2B, one unit cell is considered, with their boundaries in the $x$-$y$ plane set as periodic and other boundaries in the $z$-direction applied as hard. In Figure 2C, the left and right boundaries (along the $x$-direction) of the supercell are set as periodic, while the other boundaries are also applied as hard. In Figure 4C and 5B, scattering boundary conditions are imposed at the input and output ports, with hard boundary conditions applied in the other directions.

*Experimental measurements*

All samples are fabricated using 3D printing technology with resin, and the wall thickness is taken as 2 mm. The sound signal is scanned by an acoustic detector (MIC ⅛-inch microphone). The phases and amplitudes of the pressure fields are collected using a DAQ card (NI 9234).

For band structure measurements, two loud speakers with 0 or $\pi$ phase modulations are placed at the inputs of the MMI waveguide to generate even or odd acoustic sources, which are guided into the sample to excite forward-propagating waves. The acoustic detector probes the excited pressure fields at each hole, with all other holes sealed to prevent excessive sound wave leakage. The obtained real-space data are further processed using Fourier transformation with Matlab's built-in $fft$ function to obtain the band structures in the momentum space. The zeros padding method is employed to increase frequency resolution. The measured band structures are shown in Figure 3B and 3C using color maps.

For transmission spectrum measurements, an acoustic loud speaker is placed at the input port I1, and two acoustic detectors are set at the outputs O1 and O2 to probe the excited pressure fields. The power ratio spectra are obtained, as shown in Figure 4B and 4E. For the intensity measurements, the excitation setup is identical to that used for transmission spectrum measurements. An acoustic detector probes the pressure fields at each hole, with the remaining holes sealed to minimize sound wave leakage. The measured intensity distributions are shown in Figure 5C and 5D.

## Acknowledgements

We acknowledge support from the National Natural Science Foundation of China (Grant No. 12222407), the National Key R&D Program of China (Grants No. 2023YFA1407700 and No. 2023YFA1406904), and the Key R&D Program of Jiangsu Province (Grant No. BK20232015).

## Conflict of Interest

The authors declare no competing interests.

## Data Availability Statement

The data reported in the main text and the Supporting Information are available from the corresponding authors upon reasonable request.